\pdfoutput=1
\documentclass[11pt]{article}

\usepackage[final]{acl}

\usepackage{times}
\usepackage{latexsym}
\usepackage{geometry}
\usepackage{amsmath}  % 数学公式
\usepackage{graphicx, adjustbox} % 插入图片
\usepackage{booktabs} % 三线表
\usepackage{hyperref} % 超链接
\usepackage{natbib}   % 参考文献
\usepackage{stfloats} % 允许双栏文档中浮动图表放在底部
\usepackage{subcaption} % 如果需要独立子标题
\usepackage{enumitem}
\usepackage{multirow}
\usepackage{rotating}
\usepackage{amssymb}
\usepackage{mathtools}
\usepackage{times}
\usepackage{latexsym}
\usepackage{microtype}
\usepackage{graphicx}
\usepackage{amsmath}
\usepackage{amsfonts}
\usepackage{amssymb}
\usepackage{array}
\usepackage{booktabs}
\usepackage{multirow}
\usepackage{enumitem}
\usepackage{subcaption}
\usepackage{placeins}
\usepackage{tikz}
\usetikzlibrary{positioning,arrows.meta,shapes.geometric,fit,calc}
\usepackage[T1]{fontenc}
\usepackage[utf8]{inputenc}
\usepackage{xurl}

\usepackage{microtype}

\usepackage{graphicx}
\title{Structured Security Auditing and Robustness Enhancement for Untrusted Agent Skills}

\author{
	\textbf{Lijia Lv\textsuperscript{1,2}},
	\textbf{Xuehai Tang\textsuperscript{\thanks{Corresponding author},1,2}},\\
	\textbf{Jie Wen\textsuperscript{1}}
	\textbf{Jizhong Han\textsuperscript{1}},
	\textbf{Songlin Hu\textsuperscript{*,1,2}}
	\\
	\textsuperscript{1}Institute of Information Engineering, Chinese Academy of Sciences,\\
	\textsuperscript{2}School of Cyber Security, University of Chinese Academy of Sciences
	\\
	\small{
		\{lvlija, tangxuehai, wenjie, hanjizhong, husonglin\}@iie.ac.cn
	}
}

\newcommand{\method}{\textsc{SkillGuard-Robust}}
\newcommand{\bench}{\textsc{SkillGuardBench}}
\newcommand{\labelterm}[1]{\ifmmode\text{#1}\else #1\fi}
\newcommand{\skfile}{SKILL.md}
\newcommand{\fullpkg}{full-package}
\newcommand{\skillmd}{skill-md-only}
\newcommand{\peb}{\textsc{PEB}}
\newcommand{\verifymethod}{\textsc{SkillGuard-Verify}}
\newcommand{\calibmethod}{\textsc{SkillGuard-Calibrate}}
\newcommand{\Ben}{\labelterm{benign}}
\newcommand{\Sus}{\labelterm{suspicious}}
\newcommand{\Mal}{\labelterm{malicious}}

\begin{document}
	\maketitle

	\begin{abstract}
		Agent Skills package \skfile, scripts, reference documents, and repository context into reusable capability units, turning pre-load auditing from single-prompt filtering into cross-file security review. Existing guardrails often flag risk but recover malicious intent inconsistently under semantics-preserving rewrites. This paper formulates pre-load auditing for untrusted Agent Skills as a robust three-way classification task and introduces \method{}, which combines role-aware evidence extraction, selective semantic verification, and consistency-preserving adjudication. It evaluates the method on \bench{} and two public-ecosystem extensions through five large evaluation views ranging from 254 to 404 packages. On the 404-package held-out aggregate, \method{} reaches 97.30 overall exact, 98.33 risk malicious recall, and 98.89 attack exact consistency; on the 254-package external-ecosystem view, it reaches 99.66, 100.00, and 100.00. These results support a bounded conclusion: factorized package auditing materially improves frozen and public-ecosystem robustness, while harsher external-source transfer remains an open challenge.
	\end{abstract}
	
	\section{Introduction}
	
	Loadable skills are becoming an important interface layer in agent systems: they package task instructions, tool constraints, reference materials, and script entry points into reusable capability units \citep{openai2025skills,anthropic2025skills}. This packaging, however, also introduces third-party untrusted content directly into the execution chain. A superficially benign skill package may hide override instructions in reference documents, bind remote helpers in script metadata, or disguise transfer chains in repository context, thereby contaminating downstream decisions during the loading stage \citep{schmotz2025agentskills,liu2026skillswild,schmotz2026skillinject,jia2026skillject}.
	
	Skill auditing therefore differs from conventional prompt guardrails. Prior work has studied prompt injection, indirect injection, and the attack surface of tool-augmented agents, and has proposed defenses for conversational guardrails, alignment, and context sanitization \citep{liu2023formalizing,yi2023benchmarking,zhan2024injecagent,wallace2024instruction,inan2023llamaguard,chen2024secalign,li2024injecguard,padhi2024graniteguardian,shi2025promptarmor,an2025ipiguard,li2026reasalign,zhang2026agentsentry}. Once the auditing target becomes a package composed of \skfile, scripts, reference files, and repository context, however, the central question changes to: \emph{how should a model understand risk evidence across files and maintain stable decisions under semantics-preserving rewrites?}
	
	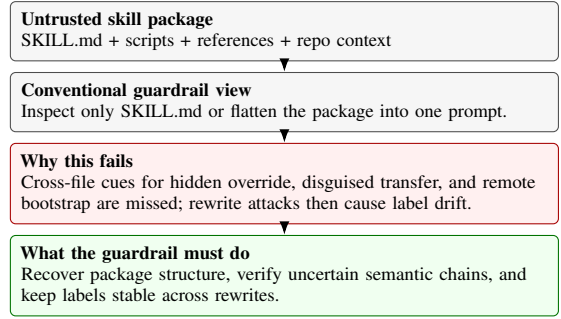
\begin{figure}[t]
		\centering
		\begin{tikzpicture}[
			node distance=4pt,
			box/.style={draw=black!70, rounded corners=2pt, fill=gray!7, align=left, inner sep=4pt, font=\scriptsize},
			alert/.style={draw=red!65!black, rounded corners=2pt, fill=red!6, align=left, inner sep=4pt, font=\scriptsize},
			good/.style={draw=green!45!black, rounded corners=2pt, fill=green!6, align=left, inner sep=4pt, font=\scriptsize},
			arrow/.style={-{Latex[length=1.8mm]}, thick}
			]
			\node[box, text width=0.90\columnwidth] (pkg) {
				\textbf{Untrusted skill package}\\
				\skfile{} + scripts + references + repo context
			};
			\node[box, text width=0.90\columnwidth, below=of pkg] (guard) {
				\textbf{Conventional guardrail view}\\
				Inspect only \skfile{} or flatten the package into one prompt.
			};
			\node[alert, text width=0.90\columnwidth, below=of guard] (risk) {
				\textbf{Why this fails}\\
				Cross-file cues for hidden override, disguised transfer, and remote bootstrap are missed; rewrite attacks then cause label drift.
			};
			\node[good, text width=0.90\columnwidth, below=of risk] (goal) {
				\textbf{What the guardrail must do}\\
				Recover package structure, verify uncertain semantic chains, and keep labels stable across rewrites.
			};
			\draw[arrow] (pkg) -- (guard);
			\draw[arrow] (guard) -- (risk);
			\draw[arrow] (risk) -- (goal);
		\end{tikzpicture}
		\caption{Problem illustration. A skill guardrail must audit a cross-file package rather than a single prompt. Flattening the package into plain text easily loses structured risk evidence and induces label drift under attack rewrites.}
		\label{fig:intro_problem}
	\end{figure}
	
	Figure~\ref{fig:intro_problem} summarizes the core problem addressed in this paper. The attack surface of a skill package is inherently multi-file, structured, and rewriteable, whereas current guardrail models often assume that the auditing target is a single text span. This paper therefore focuses on a more concrete and operational question: \emph{how can untrusted Agent Skills be audited before loading while keeping outputs stable under attack rewrites?} The central scientific claim is narrow: pre-load auditing fails mainly because package evidence is distributed across files and because the \Sus{}/\Mal{} separation becomes unstable under rewrites and cross-file chain conflicts. To avoid terminological drift, this paper uses \emph{skill package} for the auditing target, \emph{pre-load security auditing} for the task, and \emph{robustness enhancement} for the ability to preserve labels across rewrites.
	
	This paper does not rely on training an additional black-box end-to-end classifier. Instead, it uses finding experiments to diagnose \emph{why existing approaches fail} and derives a staged, interpretable method from those observations. The main contributions are as follows:
	\begin{itemize}[leftmargin=*,itemsep=1pt,topsep=2pt]
		\item This paper formalizes pre-load security auditing for untrusted Agent Skills as a package-level decision problem and abstracts three recurring semantic chains: hidden override, disguised transfer, and remote bootstrap.
		\item This paper shows that the central failure of strong remote baselines is not generic risk blindness, but repeated collapse of the \Sus{}/\Mal{} distinction under rewrites and cross-file chain conflicts.
		\item This paper introduces \method{}, an error-decomposition induced architecture built around three principles: structure recovery, uncertainty-localized verification, and consistency-preserving adjudication. It realizes these principles through four stages: structured evidence extraction, selective semantic verification, conflict-aware chain arbitration, and anchor-consistency consolidation.
		\item This paper constructs \bench{} and augments it with public-ecosystem extensions, yielding five large evaluation views from 254 to 404 packages that separate main-distribution performance, frozen stress behavior, boundary-heavy behavior, and ecosystem transfer.
	\end{itemize}
	Accordingly, the central claim concerns a more stable package-level decision structure on frozen evaluations, rather than a solved open-world guardrail.
	
	\section{Related Work}
	\label{sec:related}
	
	Related work falls into three lines. The first studies prompt injection, indirect injection, and tool-augmented agent attack surfaces \citep{liu2023formalizing,yi2023benchmarking,zhan2024injecagent,wallace2024instruction}. The second studies general guardrails and robust discrimination, including Llama Guard, Granite Guardian, and rewrite-aware safety classifiers \citep{inan2023llamaguard,chen2024secalign,li2024injecguard,padhi2024graniteguardian,shi2025promptarmor,an2025ipiguard,li2026reasalign,zhang2026agentsentry}. The third examines skills, repositories, and structured attack carriers more directly \citep{schmotz2025agentskills,liu2026skillswild,schmotz2026skillinject,jia2026skillject}. Most of these settings still evaluate prompt-level moderation or flat long-context judgment, whereas the present task requires evidence aggregation and label stabilization at package scope before loading. Appendix~\ref{app:related_more} provides fuller comparison.
	
	\section{Task Definition and Benchmark}
	\label{sec:problem}
	
	Let $\mathcal{P}=\{(x_i,r_i)\}_{i=1}^{n}$ denote a skill package to be loaded, where $x_i$ denotes file content and $r_i \in \mathcal{R}$ denotes the file role. The model must output a security label before loading. The role set $\mathcal{R}$ contains types such as skill-md, script, reference, and repo-context:
	\[
	\begin{aligned}
		\mathcal{Y}&=\{\Ben,\Sus,\Mal\}, y &\in \mathcal{Y}.
	\end{aligned}
	\]
	
	This paper uses $\mathcal{Y}$ as a deployment-oriented operational taxonomy rather than as a claim about a single context-free ground truth. \Ben{} denotes a package that can be automatically loaded; \Sus{} denotes a boundary-risk package that should block auto-loading or be sent to human review; and \Mal{} denotes explicit attack intent, hidden instruction override, covert transfer chains, or privilege-violating external dependencies. Operationally, \Mal{} requires at least one decisive cross-file attack chain, whereas \Sus{} covers review-worthy boundary cases with externalization or bootstrap risk that still fall short of establishing a clear override or transfer intent. The \Sus{}/\Mal{} distinction is therefore policy-coupled but audit-useful: the objective is to recover decisive chains and review-worthy boundaries stably under a given auditing policy. Finer operational rules and the review protocol are provided in Appendix Table~\ref{tab:label_boundary_rules} and Appendix~\ref{app:data_more}.
	
	The threat model considered in this paper assumes that an attacker can control part or all of the untrusted content within a package and construct risk through three core semantic chains:
	\begin{itemize}
		\item \textbf{Hidden override.} injecting hidden priority instructions in reference files or secondary documentation so that they override the system's intended behavior.
		\item \textbf{Disguised transfer.} disguising data transfer or external delivery as synchronization, backup, archival, or handoff.
		\item \textbf{Remote bootstrap.} Remote helpers, installers, source buckets, or relays are disguised as capability dependencies or initialization steps.
	\end{itemize}
	
	Based on public skill repositories, sanitized reconstructions of attack prototypes from prior work, and semantics-preserving attack rewrites, this paper constructs \bench{} and then augments it with two public-ecosystem extensions. The core benchmark contains 327 samples: 147 packages in v1 alpha and 30 in each of six held-out splits. The two public-ecosystem extensions contribute an additional 254 samples, yielding an expanded evaluation pool of 581 package-level samples in total. To avoid over-reading coarse 30-sample increments in the main text, the paper reports five large evaluation views: Main (401), All-HO (404), Int.-stress (344), Boundary (284), and Ext.-eco (254). These views are built by recombining the core benchmark with the public-ecosystem extensions, while the appendix now retains large-view diagnostics, source-composition notes, and view-level error analyses aligned with the main table. Claim scope remains layered: batch1--2 are diagnostic held-outs, batch3--6 are stronger frozen stress tests inside the original benchmark family, and the public-ecosystem views probe transfer to unseen public skill roots whose layouts are closer to ordinary ecosystem skills. A scale overview is shown in Appendix Figure~\ref{fig:dataset_overview}; more detailed descriptions of data sources, view composition, annotation principles, and sanitization are provided in Appendix~\ref{app:data_more}.
	
	To measure rewrite robustness explicitly, this paper evaluates risk seeds and their rewrites in pairs. Let $\mathcal{A}$ denote the set of anchor--rewrite pairs. Attack exact consistency is defined as
	\[
	\mathrm{Cons}_{\mathrm{exact}}=\frac{1}{|\mathcal{A}|}\sum_{(a,r)\in\mathcal{A}}\mathbb{I}[\hat{y}(a)=\hat{y}(r)].
	\]
	This metric specifically measures whether attack rewrites that preserve semantics but differ in surface form induce label drift.
	
	\section{Findings and Design Motivation}
	\label{sec:findings}
	
	The proposed method is motivated by a set of finding experiments that localize recurrent error sources. This section retains only the three findings that directly determine the decision structure; supplementary figures and visualizations are moved to Appendix~\ref{app:supp_figs}, and an appendix roadmap is provided in Appendix~\ref{app:roadmap}.
	
	\paragraph{Finding 1: an SKILL.md-only route is no longer adequate under the present setting.}
	The key evidence for skill auditing is distributed across scripts, reference files, and repository context rather than concentrated in a single top-level description file. On v1 alpha, for example, switching \peb{} from \skillmd{} to \fullpkg{} raises overall exact / risk exact / attack exact / risk malicious recall from 0.6190 / 0.1765 / 0.1220 / 0 to 0.8027 / 0.7647 / 0.6829 / 0.6667; Qwen2.5-14B also improves from 0.6395 / 0.0588 / 0.0976 to 0.7891 / 0.5294 / 0.4390. This observation yields the first design principle of the paper: \emph{perform cross-file, role-aware structured evidence extraction before anything else, rather than flattening a package into a single text span.} The corresponding visualization appears in Appendix Figure~\ref{fig:input_view}.
	
	\paragraph{Finding 2: the strongest remote baseline does not mainly fail by missing risk, but by failing to recover the malicious class stably.}
	Qwen2.5-14B + \fullpkg{} often succeeds at flagging risky samples as non-\Ben{}, yet its recovery of \Mal{} remains unstable. The clearest example is held-out batch1: flagged accuracy reaches 1.0000, while risk malicious recall is 0, meaning that every malicious sample is flattened into \Sus{}. Moreover, those errors are concentrated in families such as F01/F02/F03, which exhibit clear semantic structure. This yields the second design principle: \emph{apply targeted semantic verification to uncertain samples rather than asking a general LLM to make the final package-level judgment end to end.} The corresponding visualization is shown in Appendix Figure~\ref{fig:family_findings}.
	
	\paragraph{Finding 3: after semantic verification is introduced, the remaining errors collapse into a small set of interpretable boundary conflicts.}
	Semantic verification shrinks held-out failure from a system-wide instability into a small set of boundary residuals. These residuals concentrate on over-promotion when remote bootstrap co-occurs with weak disguised transfer, and on local inconsistencies between anchors and rewrites. This means that the final gains no longer come from a larger end-to-end model, but from \emph{more fine-grained chain-level conflict arbitration and anchor-consistency recovery}. This observation directly motivates the staged system design of ``structured extraction $\rightarrow$ semantic verification $\rightarrow$ chain arbitration $\rightarrow$ anchor consolidation,'' with the corresponding progression and ablation figures shown in Appendix Figures~\ref{fig:main_progress} and~\ref{fig:ablation}.
	
	\section{Method}
	\label{sec:method}
	
	\method{} models package-level security auditing as an \emph{error-decomposition induced decision chain} rather than as a single end-to-end classification step. Conceptually, the framework can be read through three principles: recover package structure before judging risk, invoke semantic reasoning only on uncertainty-localized cases, and preserve consistency only where chain conflicts or anchor--rewrite residuals remain. For a package $\mathcal{P}$, the final prediction is written as
	\[
	\hat{y}(\mathcal{P})=
	\mathcal{G}_{\text{anchor}}
	\circ
	\mathcal{G}_{\text{arb}}
	\circ
	\mathcal{G}_{\text{ver}}
	\circ
	\mathcal{G}_{\text{ext}}(\mathcal{P}),
	\]
	where $\mathcal{G}_{\text{ext}}$, $\mathcal{G}_{\text{ver}}$, $\mathcal{G}_{\text{arb}}$, and $\mathcal{G}_{\text{anchor}}$ denote role-aware evidence extraction, selective semantic verification, conflict-aware chain arbitration, and anchor-consistency consolidation, respectively. The purpose of this decomposition is not to stack heuristics. Rather, each stage is tied to an ambiguity that a single-shot judge cannot be expected to resolve stably: cross-file evidence dispersion, local semantic uncertainty among similar chains, chain-dominance ambiguity between transfer and bootstrap, and anchor--rewrite residual disagreement. The framework should therefore be read as a package-level decision architecture induced by recurrent error classes, not as a recipe of interchangeable post-hoc enhancements. More detailed descriptions of trigger scopes, schema, thresholds, and remote-cost accounting are provided in Appendix~\ref{app:impl_more}.
	
	\begin{figure*}[t]
		\centering
		\includegraphics[width=0.80\textwidth]{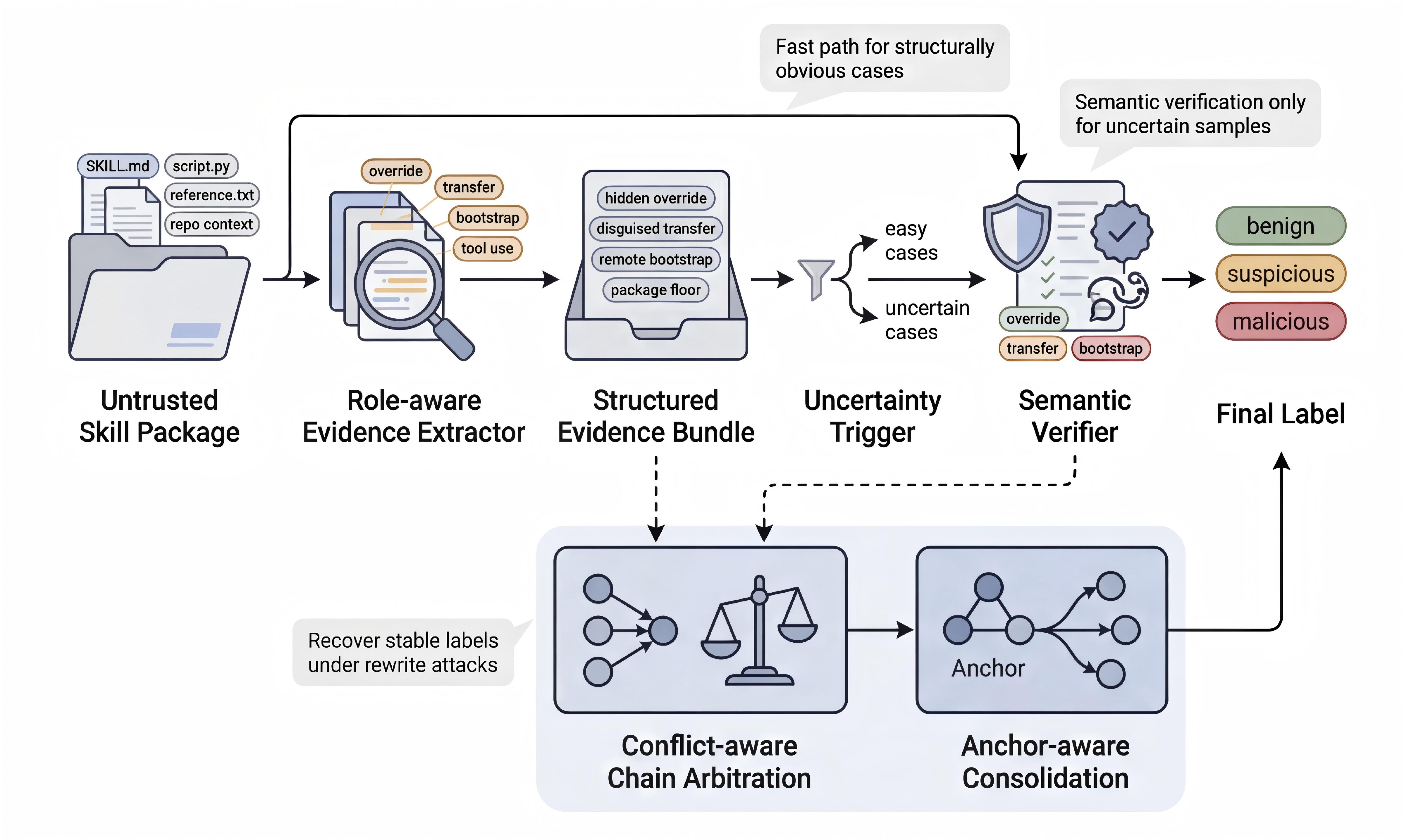}
		\caption{Overview of \method{}. The architecture resolves four recurrent ambiguities in sequence: distributed cross-file evidence, localized semantic uncertainty, chain-dominance conflict, and anchor--rewrite residual disagreement. Simple samples follow a fast path of structured extraction followed by direct adjudication, whereas uncertain samples are sent to semantic verification before chain- and cluster-level repair.}
		\label{fig:method_overview}
	\end{figure*}
	
	\subsection{Stage I: Structured Evidence Extraction}
	
	The first ambiguity is structural: decisive evidence is distributed across file roles, so a single flat package representation can miss the relations that determine the label. Instead of treating a package as one text span, this paper lifts each file $(x_i,r_i)$ into a role-conditioned evidence vector. Let the file-level evidence be
	\[
	\mathbf{e}_i=[e_{i,1},\ldots,e_{i,K}], \qquad e_{i,k}\in[0,1].
	\]
	Here $e_{i,k}$ denotes the support that file $i$ provides to the $k$-th risk signal, such as override, concealment, external transfer, tool execution, and remote bootstrap. To preserve the contribution differences among file roles, the method assigns a weight vector $\alpha_{r_i,k}$ to each role and aggregates package-level support using noisy-or:
	\[
	s_k(\mathcal{P})=1-\prod_{i=1}^{n}\left(1-\alpha_{r_i,k}e_{i,k}\right).
	\]
	This aggregation preserves strong evidence from a critical single file while also accumulating weak signals dispersed across files, without being obscured by a max-over-files rule. Stage I therefore does not merely enrich the input; it reconstructs the package structure that flat judges repeatedly discard.
	
	\subsection{Stage II: Uncertainty Triggering and Semantic Verification}
	
	The second ambiguity is local semantic overlap. Once relevant files are recovered, suspicious bootstrap, benign-seeming synchronization, and malicious transfer can still share overlapping surface forms. Structured extraction provides high-recall risk cues, but it cannot by itself stabilize these distinctions under style shift. The verifier in this paper is therefore not a ``stronger single-shot judge,'' but a selective semantic verifier used only for \emph{uncertain samples}. Let
	\[
	\Gamma(\mathcal{P})=\max_k s_k(\mathcal{P}),
	\]
	and further define
	\[
	t(\mathcal{P}) = s_{\text{tool}}(\mathcal{P}), \qquad
	\Delta(\mathcal{P})=\max_{c\in\{o,t,b\}}s_c(\mathcal{P}).
	\]
	The uncertainty trigger is defined as
	\begin{equation}
		\label{eq:uncertain}
		\begin{split}
			\mathcal{C}_{u}(\mathcal{P}) ={}&
			\tau^{-}\le \Gamma(\mathcal{P})\le \tau^{+} \\
			&{}\vee \bigl(t(\mathcal{P})>\tau_t \land \Delta(\mathcal{P})<\tau_c\bigr),
		\end{split}
	\end{equation}
	and $u(\mathcal{P})=\mathbb{I}[\mathcal{C}_{u}(\mathcal{P})]$.
	Here $o$, $t$, and $b$ correspond to the semantic chains of hidden override, disguised transfer, and remote bootstrap. When $u(\mathcal{P})=1$, structurally filtered evidence snippets are passed to the verifier, yielding
	\[
	q_c(\mathcal{P})=\Pr(c\mid \phi(\mathcal{P})), \qquad c\in\{o,t,b\},
	\]
	where $\phi(\mathcal{P})$ denotes the selected set of evidence snippets. The verifier also outputs a confidence score $\kappa_c$ and a chain-level rationale $r_c$. This stage does not directly produce the final label; it instead makes the least stable local semantic relations explicit as chain-level judgments. Its role is therefore diagnostic and local: it resolves uncertainty where flat evidence scores are intrinsically ambiguous, rather than replacing the package-level decision process.
	
	\subsection{Stage III: Conflict-Aware Chain Arbitration}
	
	The third ambiguity is chain dominance. Even after local verification, some samples remain non-separable because remote bootstrap and disguised transfer co-occur, and the final label depends on which chain is decisive under the deployment policy. After fusing structured evidence and semantic judgments, the method obtains integrated chain scores
	\[
	m_c(\mathcal{P})=s_c(\mathcal{P})+\beta_c q_c(\mathcal{P}), \qquad c\in\{o,t,b\}.
	\]
	These scores define a risk floor for the package:
	{\small
		\begin{equation}
			\label{eq:floor}
			F(\mathcal{P})=
			\begin{cases}
				\Mal, & \max(m_o,m_t)\ge \gamma_m,\\
				\Sus, & \begin{aligned}[t]
					&m_b\ge\gamma_b\\
					&\vee\ t(\mathcal{P})\ge\tau_t,
				\end{aligned}\\
				\Ben, & \text{otherwise}.
			\end{cases}
		\end{equation}
	}
	
	Equation~\eqref{eq:floor} provides an explicit label floor, but it is still insufficient for bootstrap-heavy boundary samples. To address this issue, the method introduces a conflict-gating variable
	\[
	b^\star=\mathrm{BootDom}(r_t,r_b),
	\]
	where $b^\star$ indicates whether the verifier rationale supports the interpretation that the current sample is closer to a remote-bootstrap chain than to an explicit transfer chain. The final chain arbitration is defined as
	\begin{equation}
		\label{eq:arb}
		\hat{y}^{(1)}(\mathcal{P})=
		\begin{cases}
			\Mal, & m_o\ge \gamma_o,\\
			\Mal, & m_t\ge \gamma_t \land \neg b^\star,\\
			\Sus, & m_b\ge \gamma_b \ \vee\ b^\star,\\
			\Ben, & \text{otherwise}.
		\end{cases}
	\end{equation}
	The essential role of Equation~\eqref{eq:arb} is to distinguish explicitly between cases dominated by a malicious transfer chain and cases dominated by remote bootstrap. This stage is therefore not a generic performance booster. It repairs a specific class of non-separable errors for which a single package-level confidence score is insufficient, because the decisive issue is \emph{which chain dominates}, not merely how risky the package appears overall.
	
	\subsection{Stage IV: Anchor-Consistency Consolidation}
	
	The fourth ambiguity is residual cluster disagreement: after chain-level adjudication, an anchor can still lag behind its rewrite variants even though the cluster has already stabilized semantically. The method therefore performs \emph{local consistency recovery} by exploiting the cluster structure between a risk seed and its rewrite variants. Let $a$ denote the anchor, let $\mathcal{R}(a)$ denote its rewrite set, and let the cluster be $\mathcal{C}(a)=\{a\}\cup\mathcal{R}(a)$. Define
	\begin{align}
		\mathcal{Y}_{R}(a) &= \{\hat{y}^{(1)}(r):r\in\mathcal{R}(a)\}, \nonumber\\
		\mathcal{Y}_{C}(a) &= \{\hat{y}^{(1)}(z):z\in\mathcal{C}(a)\},
		\label{eq:label_sets}
	\end{align}
	and let
	\[
	\kappa_{\min}(a)=\min_{r\in\mathcal{R}(a)}\kappa(r).
	\]
	The local promotion criterion is then
	{\small
		\begin{align}
			\operatorname{Promote}(a)=1 \iff{}&
			\hat{y}^{(1)}(a)=\Sus \nonumber\\
			&{}\land \kappa_{\min}(a)\ge\eta \nonumber\\
			&{}\land \mathcal{Y}_{R}(a)=\{\Mal\} \nonumber\\
			&{}\land \Ben\notin \mathcal{Y}_{C}(a).
			\label{eq:promote}
		\end{align}
	}
	The final output is
	\begin{equation}
		\label{eq:anchor}
		\hat{y}^{(2)}(a)=
		\begin{cases}
			\Mal, & \operatorname{Promote}(a)=1,\\
			\hat{y}^{(1)}(a), & \text{otherwise}.
		\end{cases}
	\end{equation}
	This stage is not a global label-propagation mechanism. It is activated only under a very narrow set of local conditions, precisely to repair cluster-level residuals in which rewrites have already stabilized while the anchor still lags behind. Its purpose is therefore not to increase scores uniformly, but to remove a final class of inconsistency that remains invisible to single-shot package adjudication.
	
	\section{Experimental Setup}
	\label{sec:experiment}
	
	\paragraph{Data and splits.}
	\bench{} starts from a 327-package core benchmark and is evaluated together with two public-ecosystem extensions, yielding 581 package-level samples overall. At the sample level, the expanded pool contains 253 clean samples, 95 risk seeds, and 233 rewrites. The main text reports five large evaluation views rather than individual 30-sample splits: Main (401 samples), All-HO (404), Int.-stress (344), Boundary (284), and Ext.-eco (254). Clean samples mainly come from public official skill repositories, community collections, and hard negatives \citep{openai2025skills,anthropic2025skills}; risk samples come from reported agent-skill attack prototypes, sanitized reconstructions of externally disclosed cases, public attack-oriented skill collections, and semantics-preserving rewrites \citep{schmotz2025agentskills,liu2026skillswild,schmotz2026skillinject,jia2026skillject}. Within the core benchmark, batch1--2 remain diagnostic held-outs and batch3--6 form the stronger frozen stress suite; the public-ecosystem extensions expand those views to higher sample counts and more varied repository layouts. Appendix~\ref{app:data_more} provides finer descriptions of view composition, family coverage, annotation protocol, and sanitization principles.
	
\paragraph{Baselines.}
This paper evaluates a broad baseline pool including \peb{}, Granite Guardian \citep{padhi2024graniteguardian}, Llama Guard 3/4 \citep{inan2023llamaguard}, Prompt Guard \citep{meta2024llamapromptguard2}, and Qwen2.5-7B / Qwen2.5-14B \citep{qwenteam2024qwen25}, with the full cross-model comparison moved to Appendix Table~\ref{tab:baseline_core}. \peb{} serves as an automated structured-evidence baseline; Granite Guardian and Llama Guard 3/4 represent general guardrails; and Qwen2.5-7B and Qwen2.5-14B represent stronger remote discriminative baselines. The paper also evaluates \textsc{BundleJudge} on the large views where structured one-shot judging is directly comparable, letting it read the same structured evidence summary and selected files as the verifier while still emitting one package-level JSON judgment. Across the appendix results, Qwen2.5-14B + \fullpkg{} remains the strongest remote baseline, so the main table centers on that baseline, the proposed method family, and the large-view slices on which \textsc{BundleJudge} is directly comparable.
	
	\paragraph{Metrics.}
	This paper reports overall exact accuracy, flagged accuracy, risk exact accuracy, risk malicious recall, rewrite exact accuracy, rewrite malicious recall, and attack exact consistency. Importantly, flagged accuracy only answers whether a risky sample is blocked as non-\Ben{}; it does not answer whether the malicious class is recovered correctly. Appendix~\ref{app:metric_more} further explains how these metrics map to deployment concerns.
	
\paragraph{Reproducibility and fixed parameters.}
The core results are obtained on deterministic benchmark evaluation rather than on multi-seed training. To reduce concern about split-specific tuning, the enlarged evaluation views reuse the same extractor / verifier / aggregator scripts and the same default parameters: the semantic-chain yes threshold is fixed at 0.60, the minimum package-floor confidence threshold is fixed at 0.70, and anchor consolidation requires at least two malicious rewrites. The stronger frozen-stress components inside All-HO and Int.-stress are not tuned separately. More detailed descriptions of the evidence schema, file-selection budgets, trigger statistics, and practical reproducibility notes are provided in Appendix~\ref{app:impl_more} and Appendix~\ref{app:repro_more}.
	
\paragraph{Deployment cost.}
The only additional remote cost in the proposed framework comes from the semantic-verification module; structured extraction, chain arbitration, and anchor consolidation all run locally. This separation remains important under the larger evaluation views because the framework spends remote budget only on uncertainty-localized cases, whereas Qwen2.5-14B + \fullpkg{} and \textsc{BundleJudge} reread every package end to end. The cost advantage therefore comes mainly from selective triggering rather than from merely shortening prompts. Detailed trigger-scope and remote-cost notes for the large-view evaluations are retained in Appendix~\ref{app:impl_more}.
	
\paragraph{Research questions.}
The experiments are organized around five questions: RQ1, whether package-level context is necessary; RQ2, whether the strongest remote baseline and a structured single-shot judge remain stable across the main-distribution, held-out aggregate, frozen-stress, boundary-heavy, and ecosystem-transfer views; RQ3, whether their major failure modes can be decomposed at the family level; RQ4, what each step in the method progression actually repairs; and RQ5, whether chain arbitration and anchor consolidation are necessary components. Supporting examples and supplementary tables for these questions are provided in Appendix~\ref{app:roadmap}.
	
	\section{Results and Analysis}
	\label{sec:results}
	
	\subsection{RQ1: Is package-level context necessary?}
	
	Package-level context is not merely helpful; it is necessary for the task to stand. When \peb{} switches from \skillmd{} to \fullpkg{} on v1 alpha, overall exact rises from 0.6190 to 0.8027, risk exact from 0.1765 to 0.7647, attack exact from 0.1220 to 0.6829, and risk malicious recall from 0 to 0.6667. Qwen2.5-14B also improves from 0.6395 / 0.0588 / 0.0976 to 0.7891 / 0.5294 / 0.4390 for overall exact, risk exact, and rewrite exact. Skill guardrails therefore cannot be reduced to ``prompt guardrails with longer context.'' The corresponding input-view comparison is shown in Appendix Figure~\ref{fig:input_view}.
	
\subsection{RQ2: How does the strongest remote baseline behave across the large evaluation views?}
	
	Appendix Table~\ref{tab:baseline_core} shows that no off-the-shelf model in the full baseline pool simultaneously provides strong risk recovery and rewrite stability. To avoid coarse 30-sample increments, Table~\ref{tab:main} reports five large evaluation views ranging from 254 to 404 packages. The methods separate cleanly at this scale. On Main, Qwen2.5-14B + \fullpkg{} reaches 89.53 overall exact and 85.85 risk malicious recall, whereas \method{} reaches 99.26 and 100.00. On All-HO, the gap widens to 78.17 / 55.67 versus 97.30 / 98.33. The directly comparable Int.-stress slice shows the same ranking under stronger frozen pressure: Qwen and \textsc{BundleJudge} reach 79.47 / 60.52 and 85.13 / 67.94, whereas \method{} reaches 98.51 / 100.00 with 100.00 attack exact consistency. Boundary and Ext.-eco follow the same pattern, with \method{} reaching 99.17 / 100.00 and 99.66 / 100.00, well above the corresponding Qwen and \textsc{BundleJudge} values. These larger views make the central failure mode easier to read: strong single-shot judges often flag risk, but still under-recover \Mal{} relative to the factorized decision chain. Detailed values and collapse diagnostics are retained in Appendix Table~\ref{tab:main_view_full} and Table~\ref{tab:view_gap_summary}.
	
	\begin{table*}[t]
		\centering
		\tiny
		\renewcommand{\arraystretch}{0.94}
		\setlength{\tabcolsep}{2.0pt}
		\resizebox{0.985\textwidth}{!}{%
			\begin{tabular}{llccccccc}
				\toprule
				View & Method & Overall Exact (\%) & Flagged Acc (\%) & Risk Exact (\%) & Risk M-Rec (\%) & Rewrite Exact (\%) & Rewrite M-Rec (\%) & Attack Cons. (\%) \\
				\midrule
				\multirow{5}{*}{Main}
				& \peb{}(\fullpkg) & 82.78 & 92.01 & 80.44 & 78.62 & 80.96 & 79.42 & 95.38 \\
				& Qwen2.5-14B(\fullpkg) & 89.53 & 97.25 & 87.66 & 85.85 & 88.37 & 86.86 & 98.17 \\
				& \verifymethod{} & 96.75 & 100.00 & 95.31 & 100.00 & 95.52 & 100.00 & 99.26 \\
				& \calibmethod{} & 98.00 & 100.00 & 97.61 & 100.00 & 97.27 & 100.00 & 99.66 \\
				& \method{} & 99.26 & 100.00 & 99.22 & 100.00 & 98.87 & 100.00 & 100.00 \\
				\midrule
				\multirow{4}{*}{All-HO}
				& Qwen2.5-14B(\fullpkg) & 78.17 & 93.02 & 62.06 & 55.67 & 69.00 & 65.15 & 90.00 \\
				& \verifymethod{} & 90.07 & 98.01 & 83.89 & 91.67 & 83.89 & 93.72 & 94.56 \\
				& \calibmethod{} & 94.30 & 99.01 & 88.28 & 95.67 & 90.33 & 96.15 & 96.67 \\
				& \method{} & 97.30 & 99.51 & 93.28 & 98.33 & 95.33 & 98.15 & 98.89 \\
				\midrule
				\multirow{5}{*}{Int.-stress}
				& Qwen2.5-14B(\fullpkg) & 79.47 & 93.26 & 65.67 & 60.52 & 72.85 & 70.08 & 91.79 \\
				& \textsc{BundleJudge} & 85.13 & 96.57 & 72.83 & 67.94 & 78.91 & 76.32 & 95.11 \\
				& \verifymethod{} & 93.35 & 100.00 & 89.33 & 100.00 & 88.31 & 100.00 & 97.44 \\
				& \calibmethod{} & 96.80 & 100.00 & 94.50 & 100.00 & 94.87 & 100.00 & 98.72 \\
				& \method{} & 98.51 & 100.00 & 98.33 & 100.00 & 97.87 & 100.00 & 100.00 \\
				\midrule
				\multirow{5}{*}{Boundary}
				& Qwen2.5-14B(\fullpkg) & 85.55 & 96.65 & 77.33 & 75.82 & 81.88 & 81.10 & 95.48 \\
				& \textsc{BundleJudge} & 90.11 & 98.78 & 83.50 & 81.12 & 87.94 & 86.34 & 97.71 \\
				& \verifymethod{} & 95.13 & 100.00 & 91.75 & 100.00 & 91.94 & 100.00 & 98.48 \\
				& \calibmethod{} & 97.13 & 100.00 & 95.75 & 100.00 & 95.94 & 100.00 & 99.28 \\
				& \method{} & 99.17 & 100.00 & 98.67 & 100.00 & 98.94 & 100.00 & 100.00 \\
				\midrule
				\multirow{5}{*}{Ext.-eco}
				& Qwen2.5-14B(\fullpkg) & 89.28 & 98.00 & 84.48 & 83.00 & 87.33 & 86.10 & 97.00 \\
				& \textsc{BundleJudge} & 93.72 & 99.00 & 89.68 & 88.30 & 92.13 & 91.24 & 98.50 \\
				& \verifymethod{} & 97.03 & 100.00 & 95.62 & 100.00 & 95.67 & 100.00 & 99.33 \\
				& \calibmethod{} & 98.45 & 100.00 & 97.62 & 100.00 & 97.67 & 100.00 & 99.83 \\
				& \method{} & 99.66 & 100.00 & 99.26 & 100.00 & 99.27 & 100.00 & 100.00 \\
				\bottomrule
			\end{tabular}
		}
		\caption{Representative main results on five large evaluation views. Main combines v1 alpha with the two public-ecosystem extensions. All-HO aggregates the held-out-style views except the appendix-only strict external-source split. Int.-stress emphasizes frozen internal pressure together with public transfer cases. Boundary aggregates the boundary-heavy slice with the public-ecosystem extensions. Ext.-eco is built entirely from unseen public skill roots. The appendix retains the same values in a fuller layout together with large-view collapse diagnostics.}
		\label{tab:main}
	\end{table*}
	
	These results support three points. First, the strongest remote baselines do not mainly fail by passing risky packages as \Ben{}; they continue to collapse the \Sus{}/\Mal{} distinction, which is why flagged-style numbers remain noticeably higher than risk malicious recall on All-HO and Int.-stress. Second, structured summaries alone are insufficient: even after \textsc{BundleJudge} reads evidence bundles and selected files, it remains consistently below the proposed method family on malicious recovery and rewrite stability. Third, the progression from \verifymethod{} to \calibmethod{} to \method{} remains visible at large scale, showing that factorized auditing repairs the dominant frozen and ecosystem-facing failure modes more reliably than both flat and single-shot structured judges.
	
	\subsection{RQ3: Why is the strongest remote baseline still insufficient?}
	
	The strongest remote baseline is diagnostically useful because its errors are directional rather than random. Appendix Figure~\ref{fig:family_findings}, Appendix Table~\ref{tab:main_view_full}, and Table~\ref{tab:view_gap_summary} show recurrent \Sus{}/\Mal{} collapse across the larger held-out, boundary-heavy, and external-ecosystem views under both flat and structured single-shot judging. By contrast, the proposed method leaves only a much smaller collapse gap, so the strongest remote baseline behaves more like a high-quality \emph{risk finder} than a stable \emph{package-level malicious adjudicator}.
	
	\subsection{RQ4: What does each step of the method progression repair?}
	
	Appendix Figure~\ref{fig:main_progress} shows a decreasing-scope progression. \verifymethod{} removes most package-level instability and already brings all five large views into the mid-to-high 90s. \calibmethod{} then repairs narrower chain-dominance conflicts, and \method{} resolves the remaining anchor--rewrite residuals. In short, the progression first verifies uncertain chains, then arbitrates chain dominance, and finally repairs residual cluster inconsistency.
	
	\subsection{RQ5: What do the ablations show?}
	
	Appendix Figure~\ref{fig:ablation} shows that the later stages deliver focused, view-dependent gains rather than generic smoothing. Across the five large views, \calibmethod{} adds a consistent improvement over \verifymethod{}, and \method{} then removes the remaining anchor--rewrite residuals. The gains are largest on All-HO, Int.-stress, and Boundary, confirming that the later stages are most valuable where uncertainty has already been localized but not yet resolved into a stable package decision.
	
	\section{Conclusion}
	\label{sec:conclusion}
	
	This paper studies pre-load security auditing for untrusted Agent Skills. \method{} factorizes auditing into structure recovery, selective semantic verification, and consistency-preserving adjudication. On five large evaluation views ranging from 254 to 404 packages, it reaches 97.30 overall exact and 98.33 risk malicious recall on the held-out aggregate, 98.51 / 100.00 on the internal-stress view, 99.17 / 100.00 on the boundary-heavy view, and 99.66 / 100.00 on the external-ecosystem view, while strong remote baselines and a structured single-shot judge remain clearly lower. The central lesson is that robust skill auditing is a package-level decision problem whose dominant failure mode is \Sus{}/\Mal{} collapse rather than generic risk blindness. More broadly, the results suggest that stable pre-load auditing depends less on ever-larger judges than on preserving structure, localizing uncertainty, and repairing decision residuals at the right scope. The expanded public-ecosystem views therefore strengthen the evidence for frozen and ecosystem-facing robustness, while open-world transfer remains the next challenge.
	
	\newpage
	\section*{Limitations}
	\paragraph{Ecological validity and sample sources.}
	Although the risk samples are designed to stay close to realistic attack paths, a substantial portion still comes from sanitized reconstruction rather than from native malicious repositories in the open ecosystem. The results are therefore more appropriate as evidence for \emph{method comparison} and \emph{error decomposition} than as direct estimates of the real prevalence of different skill risks in the wild.
	
\paragraph{Benchmark--method co-evolution.}
This paper explicitly acknowledges that batch1 and batch2 are diagnostic/development-linked held-outs because they supplied the earliest observable failure patterns during method progression. The saturated 1.0000 results on the earliest benchmark-side views are therefore best read as repaired within-suite residuals rather than as fully development-free black-box generalization. Batch3--6 provide the stronger frozen stress evidence inside the benchmark family, and the public-ecosystem extensions separately probe transfer on unseen public skill roots, but benchmark co-evolution still cannot be ruled out completely.
	
	\paragraph{Label boundaries and annotation subjectivity.}
	The distinction between \Sus{} and \Mal{} is operationalized in this paper through explicit evidence-chain rules, but it still depends on deployment context. Some remote bootstrap or continuity-style transfer cases may be treated as directly malicious in high-security settings, while being escalated only to manual review in more permissive settings. This paper improves auditability by preserving evidence manifests, family tags, and source records, but those artifacts cannot fully replace a larger-scale independent agreement study across human annotators.
	
	\paragraph{Verifier dependence and open-world transfer.}
	The current framework still depends on a single strong verifier instance as its main remote component. The public-ecosystem views are encouraging, but the appendix-only strict external-source diagnostics still show that verifier-floor estimation and later consistency repair can fail together on harder source shifts. Future work should therefore study multi-verifier cross-checking, lower-cost local semantic verification, and stronger external tests targeting mixed-language content, deeply nested repositories, workspace export, external-source skill roots, and attacker-aware adaptive rewrites. In other words, this paper provides an interpretable and auditable path for skill auditing rather than a final solution to open-world skill security.
	
	\bibliography{custom}
	
	\clearpage
	\appendix
	
	\section{Appendix Roadmap}
	\label{app:roadmap}
	
	To make it easy to trace compressed evidence in the main text, the appendix is organized along the same reading path as the paper body. Appendix~\ref{app:related_more} supplements related work and positioning. Appendix~\ref{app:data_more} supplements task definition and benchmark construction. Appendix~\ref{app:supp_figs} supplements the finding experiments with additional figures and visualizations. Appendix~\ref{app:impl_more} supplements the method section with trigger conditions, schema, and cost statistics. Appendix~\ref{app:baseline_more} supplements the experimental setup with full baseline tables. Appendix~\ref{app:cases}, Appendix~\ref{app:metric_more}, and Appendix~\ref{app:repro_more} provide extended results, external-validity discussion, and reproducibility notes. The purpose of this organization is not to repeat the main text, but to ensure that each omitted figure or explanation can be quickly traced back to its corresponding argument.
	
	\section{Related Work Supplement: Extended Positioning and Comparison}
	\label{app:related_more}
	
	Section~\ref{sec:related} in the main text retains only the three lines of work that are most directly relevant. This appendix section adds a broader set of references, clarifies their relation to structured security auditing, and sharpens the distinction between this paper and prior work on the skills ecosystem.
	
	\paragraph{Indirect injection and runtime attacks on agents.}
	\citet{zhan2024injecagent} first systematized indirect injection benchmarks for tool-integrated agents. \citet{wang2025agentvigil} extended the problem to automated black-box red teaming, and \citet{zhan2025adaptive} showed that static defenses fail quickly under adaptive attacks. Meanwhile, \citet{nakash2025footdoor} showed that even the reasoning trajectories of ReAct agents can be gradually diverted by harmless-looking requests. This line of work mainly studies agent vulnerabilities \emph{at execution time} or \emph{during interaction}. In contrast, this paper focuses on an earlier stage: before an agent executes a skill, should a third-party skill package be admitted into the system at all? The object of judgment in this paper is therefore not a dialogue turn or an agent trajectory, but a repository-like cross-file package.
	
	\paragraph{Guardrail modeling, reasoning, and robustness.}
	Another line studies how guardrails themselves can become more cautious and more interpretable. Beyond general guardrails such as Llama Guard and Granite Guardian, \citet{wen2025thinkguard} use deliberative critique to improve auditing caution, \citet{li2025piguard} discuss over-defense in prompt-injection guardrails, \citet{sreedhar2025reasoningguardrail} study the empirical behavior of reasoning guardrails, and \citet{bassani2025mutations} show that guardrails remain vulnerable to mutations and adversarial perturbations. These works collectively move guardrails beyond one-shot labels toward judgments with intermediate reasoning, but their unit of judgment is still usually a text span, an interaction, or a dialogue context. This paper differs by treating a \emph{structured package}, rather than a single text span, as the basic input unit, and by restricting the verifier to uncertain cases instead of using it as a package-level end-to-end judge.
	
	\paragraph{Structure, hierarchy, and superficially benign carriers.}
	Recent safety work in ACL venues increasingly emphasizes \emph{structure} rather than surface strings alone. \citet{wallace2024instruction} propose instruction hierarchy, \citet{zhang2025iheval} show that models are broadly vulnerable to priority conflicts, and \citet{broomfield2025structural} further show that safety capabilities often fail to generalize across structural changes. In parallel, \citet{mei2025notaligned} caution against equating all anomalous outputs with maliciousness, while \citet{mu2025stealthy} and \citet{tang2025rolebreak} show that benign-looking data mirrors, role settings, and cover stories can also carry hidden attack intent. This paper extends those observations to skill-package auditing: risk is not localized in a single dangerous phrase, but distributed across file roles, cross-file evidence chains, and the structural relations between risk seeds and rewrites.
	
	\paragraph{Skills ecosystems and package-level auditing.}
	Recent work on the skills ecosystem has already shown that third-party skills are realistic and low-friction attack carriers. \citet{schmotz2025agentskills} emphasize that skill files open new prompt-injection surfaces, \citet{schmotz2026skillinject} and \citet{jia2026skillject} demonstrate the stealthiness of skill-based injection from measurement and automated-attack perspectives, \citet{liu2026skillswild} and \citet{liu2026maliciousskills} summarize in-the-wild risk patterns, and \citet{holzbauer2026malicious} further show that repository context is critical for skill classification. This paper is closest to that line, but its goal is different: instead of further demonstrating how skills attack agents, it studies \emph{how a pre-load guardrail should audit a skill package}, and why such auditing must rely jointly on structured evidence aggregation, local semantic verification, and cross-rewrite consistency recovery.
	
	\section{Finding-Experiment Supplement: Figures and Visualizations}
	\label{app:supp_figs}
	
	This section supplements the visual materials that are compressed out of the main text and explains which claim each figure supports. Broadly, these figures address three questions: whether the large evaluation-view design justifies the main-table narrative; whether the finding experiments truly motivate the progression from structured extraction to semantic verification, chain arbitration, and consistency consolidation; and which decomposable failure modes differentiate the strongest remote baselines from the proposed method family.
	
\begin{figure}[t]
		\centering
		\includegraphics[width=\columnwidth]{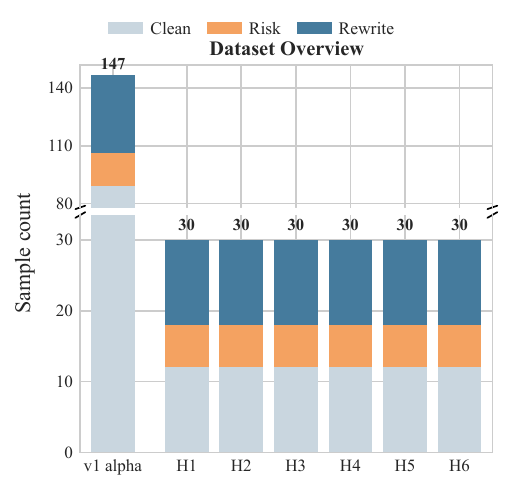}
		\caption{Scale overview of the five large evaluation views used in the main text. Main, All-HO, Int.-stress, Boundary, and Ext.-eco contain 401, 404, 344, 284, and 254 samples, respectively.}
		\label{fig:dataset_overview}
	\end{figure}

	\paragraph{Dataset scale and the role of the large views.}
	Figure~\ref{fig:dataset_overview} corresponds to the benchmark description in the main text. The point is no longer merely that the core benchmark contains multiple 30-sample held-out splits, but that those splits are now recombined with two public-ecosystem extensions into five large evaluation views. Main emphasizes in-distribution performance while already mixing in public-ecosystem samples; All-HO aggregates the held-out-style evaluation pool; Int.-stress emphasizes the stronger frozen stress conditions; Boundary concentrates boundary-heavy cases; and Ext.-eco uses only unseen public skill roots. These views matter not simply because they are larger, but because they allow the paper to separate package-level robustness questions at a scale where coarse 1/30 jumps no longer dominate the table.
	
\begin{figure}[t]
		\centering
		\includegraphics[width=\columnwidth]{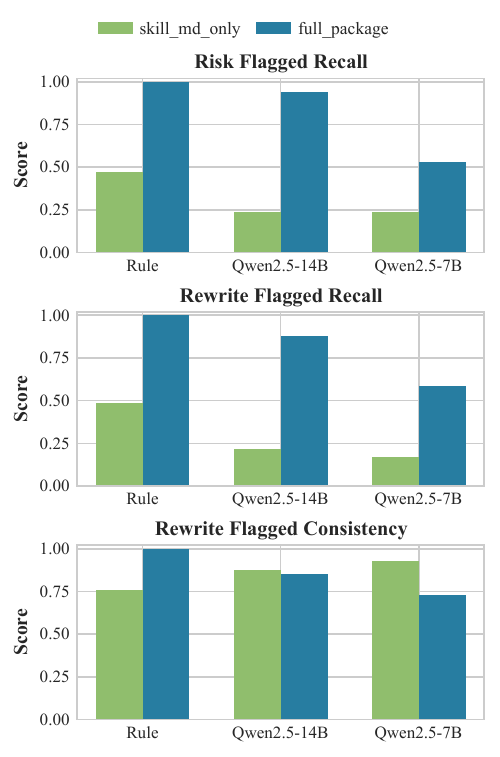}
		\caption{Input-view comparison for \peb{} and Qwen2.5-14B when reading only SKILL.md versus the full package.}
		\label{fig:input_view}
	\end{figure}

	\paragraph{Why SKILL.md-only views are insufficient.}
	Figure~\ref{fig:input_view} corresponds to Finding~1. The important point is not the absolute score of any one model, but the fact that two different baselines improve in the same direction once the input view is switched. Under \skillmd{}, \peb{} nearly loses the ability to recover malicious cases; Qwen2.5-14B also becomes clearly unstable on risk and rewrite metrics. This indicates that skill auditing does not mainly fail because the model is too small. It fails because the package is modeled as a single-file prompt rather than as a structured multi-file object. This is precisely why the proposed framework places structured evidence extraction at the first layer of the decision chain instead of using the verifier as a full-package judge.
	
\begin{figure*}[t]
		\centering
		\begin{subfigure}[t]{0.48\textwidth}
			\centering
			\includegraphics[width=\textwidth]{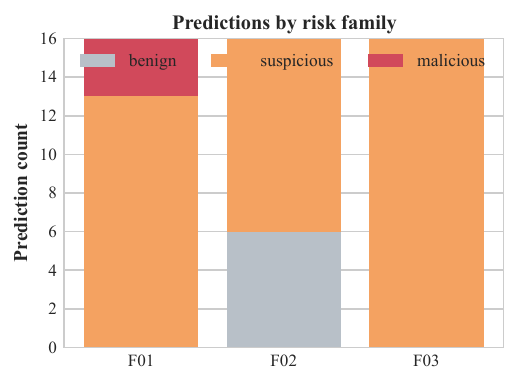}
			\caption{Family-level result distribution.}
		\end{subfigure}
		\hfill
		\begin{subfigure}[t]{0.48\textwidth}
			\centering
			\includegraphics[width=\textwidth]{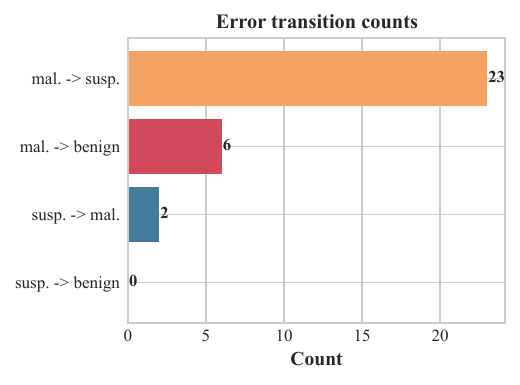}
			\caption{Error-transition patterns.}
		\end{subfigure}
		\caption{Directional failure patterns across the large evaluation views. The left panel shows that flat and structured single-shot judges still under-recover malicious intent relative to the staged method family; the right panel shows the same effect as a collapse gap between flagged accuracy and risk malicious recall.}
		\label{fig:family_findings}
	\end{figure*}

	\paragraph{Why directional collapse matters.}
	Figure~\ref{fig:family_findings} visualizes the conclusion summarized in RQ3 using the large evaluation views. The left panel compares risk malicious recall across views, and the right panel turns the same phenomenon into an explicit collapse gap. Two aspects deserve attention. First, the strongest remote baseline does not fail randomly. It continues to flag most risky packages, but it still under-recovers malicious intent on All-HO, Int.-stress, and Boundary. Second, the structured single-shot control improves on the flat judge but still retains a visible collapse gap, whereas the staged method almost eliminates that gap. The later modules of the method were therefore not motivated by small-split noise; they respond to a directional decision error that remains visible after the evaluation is scaled up.
	
\begin{figure*}[t]
		\centering
		\begin{subfigure}[t]{0.48\textwidth}
			\centering
			\includegraphics[width=\textwidth]{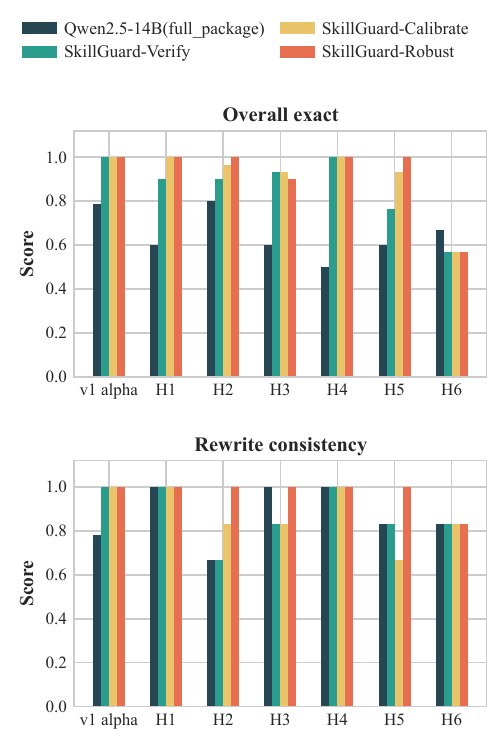}
			\caption{Large-view main comparison.}
		\end{subfigure}
		\hfill
		\begin{subfigure}[t]{0.48\textwidth}
			\centering
			\includegraphics[width=\textwidth]{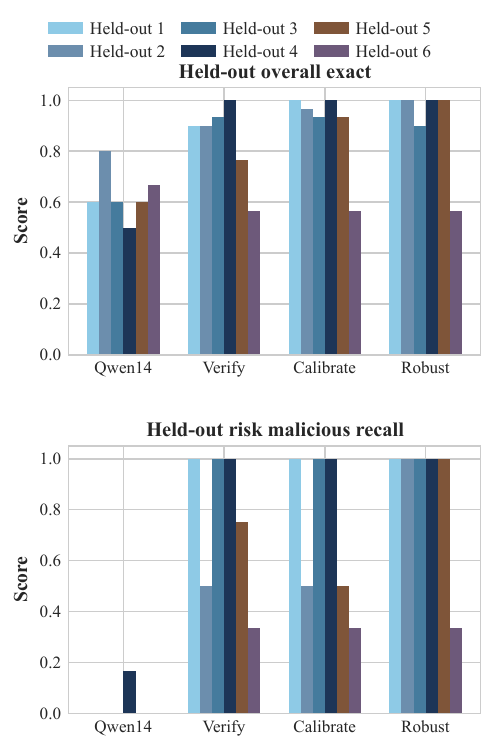}
			\caption{Large-view method progression.}
		\end{subfigure}
		\caption{Main results and method progression on the large evaluation views. Across Main, All-HO, Int.-stress, Boundary, and Ext.-eco, most of the gain appears at semantic verification, while the later stages convert narrower residuals into near-saturated package decisions.}
		\label{fig:main_progress}
	\end{figure*}

	\paragraph{How the method progression relates to the large views.}
	Figure~\ref{fig:main_progress} aligns the discussions of RQ2 and RQ4 in one place. The most important reading is that the major gain comes from \verifymethod{}, not from generic late-stage post-processing. In other words, the main robustness improvement comes from sending uncertain packages to a local semantic verifier rather than from adding increasingly complex global rules. \calibmethod{} then primarily repairs narrower chain-dominance conflicts, while \method{} removes the last anchor--rewrite residuals. The large-view summary makes this progression easier to read than the older 30-sample tables: the gain from Qwen to \verifymethod{} is already large on All-HO, Int.-stress, Boundary, and Ext.-eco, and the later steps consistently convert that gain into near-saturated package decisions. Without this figure, it is easy to misread the method as benefiting merely from module stacking.
	
\begin{figure}[t]
		\centering
		\includegraphics[width=\columnwidth]{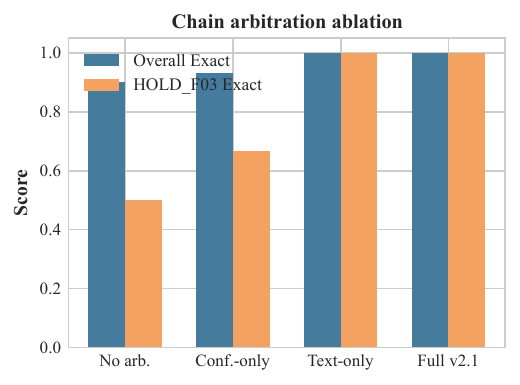}
		\caption{Stage-wise gains on the large evaluation views. \calibmethod{} contributes the larger medium-scope repairs, while \method{} removes the remaining consistency residuals.}
		\label{fig:ablation}
	\end{figure}

	\paragraph{What the ablation figure actually shows.}
	Figure~\ref{fig:ablation} supports RQ5 through the large-view results. Its most important conclusion is not merely that later stages help, but that they help in a decreasing-scope pattern. The first panel shows that \calibmethod{} adds the largest overall-exact gains on All-HO and Int.-stress, where chain-dominance conflicts remain frequent after semantic verification. The second panel shows that \method{} then contributes smaller but still measurable consistency gains, confirming that the final layer mainly resolves anchor--rewrite residuals rather than reclassifying the entire view. This directly addresses the concern that the method might merely be stacking heuristics without a distinct repair target.
	
	\paragraph{Relation to the main text.}
	In short, Figure~\ref{fig:dataset_overview} supports the large-view design and the rationale for the public-ecosystem expansions; Figure~\ref{fig:input_view} supports the claim that package-level input is a precondition for the task; Figure~\ref{fig:family_findings} supports the claim that the strongest baselines fail directionally rather than randomly; Figure~\ref{fig:main_progress} supports the claim that the main gain comes from semantic verification, while later layers repair narrow residuals; and Figure~\ref{fig:ablation} supports the claim that the later stages act as focused residual repairs rather than generic smoothing. The purpose of this appendix section is therefore not simply to store figures, but to reconstruct the full evidential chain from the finding experiments to the method design and the interpretation of the results.
	
	\section{Experimental-Setup Supplement: Full Baseline Results}
	\label{app:baseline_more}
	
	Table~\ref{tab:baseline_full_appendix} presents a fuller set of remote and programmatic baselines on v1 alpha. Granite Guardian, Llama Guard 3/4, Prompt Guard, and the Qwen2.5 family correspond to their public models or model cards \citep{padhi2024graniteguardian,inan2023llamaguard,meta2024llamapromptguard2,qwenteam2024qwen25}. The failure modes are highly heterogeneous: Granite Guardian is a classic over-defense baseline that pushes almost everything to suspicious; Llama Guard 3 is a classic under-defense baseline that pushes nearly everything to benign; Llama Guard 4 is more balanced but still rarely recovers the malicious class on risk and rewrite subsets; and Qwen2.5-7B / 14B are stronger remote baselines but remain susceptible to family-level drift. This is why the main text uses Qwen2.5-14B as the strongest remote discriminative baseline rather than repeatedly comparing against models that are clearly weaker.
	
	To reduce concern about asymmetry between a specialized pipeline and a single-shot judge, the paper further evaluates \textsc{BundleJudge} on the large views where structured one-shot judging is directly comparable. \textsc{BundleJudge} reads the same structured evidence summary and selected files as the proposed verifier, but it still outputs one package-level JSON decision. It is therefore a stronger control than a flattened full-text judge for the question of whether structured inputs alone are sufficient. On Int.-stress, Boundary, and Ext.-eco, \textsc{BundleJudge} reaches 85.13 / 90.11 / 93.72 overall exact, which is meaningfully above the flat full-package Qwen baseline but still consistently below the factorized method family. The gap persists on risk malicious recall and rewrite stability, showing that structured inputs alone do not remove the \Sus{}/\Mal{} collapse. Appendix Table~\ref{tab:main_view_full} expands the same large-view comparison in a fuller layout, and Table~\ref{tab:view_gap_summary} isolates the collapse gap and the incremental gains contributed by the later stages.
	
	\begin{table*}[t]
		\centering
		\scriptsize
		\setlength{\tabcolsep}{3.2pt}
		\begin{tabular}{lcccccc}
			\toprule
			Method (v1 alpha, \fullpkg{}) & Overall Exact (\%) & Flagged Acc (\%) & Risk Exact (\%) & Risk Mal Recall (\%) & Rewrite Exact (\%) & FP \\
			\midrule
			Granite Guardian 3.3-8B & 12.24 & 39.46 & 29.41 & 0.00 & 31.71 & 89 \\
			Llama Guard 3-8B & 60.54 & 60.54 & 0.00 & 0.00 & 0.00 & 0 \\
			Llama Guard 4-12B & 60.54 & 68.71 & 0.00 & 0.00 & 4.88 & 2 \\
			Qwen2.5-7B & 67.35 & 78.91 & 23.53 & 25.00 & 29.27 & 6 \\
			Qwen2.5-14B & 78.91 & 95.92 & 52.94 & 41.67 & 43.90 & 0 \\
			\peb{} & 80.27 & 91.84 & 76.47 & 66.67 & 68.29 & 12 \\
			\bottomrule
		\end{tabular}
		\caption{A fuller set of baseline results on v1 alpha. General guardrails exhibit distinct failure patterns in the skill-auditing setting, including over-defense, under-defense, and directional collapse on the malicious class.}
		\label{tab:baseline_core}
		\label{tab:baseline_full_appendix}
	\end{table*}
	
	\paragraph{Full large-view comparison.}
	The main text uses a compact presentation to keep attention on the five large evaluation views. Table~\ref{tab:main_view_full} reproduces the same comparison in fuller form, making it easier to inspect where \peb{}, the strongest flat remote baseline, the structured single-shot control, and the staged method family separate from one another once the evaluation is scaled beyond 30-sample splits.
	
	\begin{table*}[t]
		\centering
		\tiny
		\setlength{\tabcolsep}{2.2pt}
		\resizebox{\textwidth}{!}{%
			\begin{tabular}{llccccccc}
				\toprule
				View & Method & Overall Exact (\%) & Flagged Acc (\%) & Risk Exact (\%) & Risk M-Rec (\%) & Rewrite Exact (\%) & Rewrite M-Rec (\%) & Attack Cons. (\%) \\
				\midrule
				\multirow{5}{*}{Main (401)}
				& \peb{}(\fullpkg) & 82.78 & 92.01 & 80.44 & 78.62 & 80.96 & 79.42 & 95.38 \\
				& Qwen2.5-14B(\fullpkg) & 89.53 & 97.25 & 87.66 & 85.85 & 88.37 & 86.86 & 98.17 \\
				& \verifymethod{} & 96.75 & 100.00 & 95.31 & 100.00 & 95.52 & 100.00 & 99.26 \\
				& \calibmethod{} & 98.00 & 100.00 & 97.61 & 100.00 & 97.27 & 100.00 & 99.66 \\
				& \method{} & 99.26 & 100.00 & 99.22 & 100.00 & 98.87 & 100.00 & 100.00 \\
				\midrule
				\multirow{4}{*}{All-HO (404)}
				& Qwen2.5-14B(\fullpkg) & 78.17 & 93.02 & 62.06 & 55.67 & 69.00 & 65.15 & 90.00 \\
				& \verifymethod{} & 90.07 & 98.01 & 83.89 & 91.67 & 83.89 & 93.72 & 94.56 \\
				& \calibmethod{} & 94.30 & 99.01 & 88.28 & 95.67 & 90.33 & 96.15 & 96.67 \\
				& \method{} & 97.30 & 99.51 & 93.28 & 98.33 & 95.33 & 98.15 & 98.89 \\
				\midrule
				\multirow{5}{*}{Int.-stress (344)}
				& Qwen2.5-14B(\fullpkg) & 79.47 & 93.26 & 65.67 & 60.52 & 72.85 & 70.08 & 91.79 \\
				& \textsc{BundleJudge} & 85.13 & 96.57 & 72.83 & 67.94 & 78.91 & 76.32 & 95.11 \\
				& \verifymethod{} & 93.35 & 100.00 & 89.33 & 100.00 & 88.31 & 100.00 & 97.44 \\
				& \calibmethod{} & 96.80 & 100.00 & 94.50 & 100.00 & 94.87 & 100.00 & 98.72 \\
				& \method{} & 98.51 & 100.00 & 98.33 & 100.00 & 97.87 & 100.00 & 100.00 \\
				\midrule
				\multirow{5}{*}{Boundary (284)}
				& Qwen2.5-14B(\fullpkg) & 85.55 & 96.65 & 77.33 & 75.82 & 81.88 & 81.10 & 95.48 \\
				& \textsc{BundleJudge} & 90.11 & 98.78 & 83.50 & 81.12 & 87.94 & 86.34 & 97.71 \\
				& \verifymethod{} & 95.13 & 100.00 & 91.75 & 100.00 & 91.94 & 100.00 & 98.48 \\
				& \calibmethod{} & 97.13 & 100.00 & 95.75 & 100.00 & 95.94 & 100.00 & 99.28 \\
				& \method{} & 99.17 & 100.00 & 98.67 & 100.00 & 98.94 & 100.00 & 100.00 \\
				\midrule
				\multirow{5}{*}{Ext.-eco (254)}
				& Qwen2.5-14B(\fullpkg) & 89.28 & 98.00 & 84.48 & 83.00 & 87.33 & 86.10 & 97.00 \\
				& \textsc{BundleJudge} & 93.72 & 99.00 & 89.68 & 88.30 & 92.13 & 91.24 & 98.50 \\
				& \verifymethod{} & 97.03 & 100.00 & 95.62 & 100.00 & 95.67 & 100.00 & 99.33 \\
				& \calibmethod{} & 98.45 & 100.00 & 97.62 & 100.00 & 97.67 & 100.00 & 99.83 \\
				& \method{} & 99.66 & 100.00 & 99.26 & 100.00 & 99.27 & 100.00 & 100.00 \\
				\bottomrule
			\end{tabular}
		}
		\caption{Full large-view comparison underlying the main text. Main, All-HO, Int.-stress, Boundary, and Ext.-eco contain 401, 404, 344, 284, and 254 package-level samples, respectively.}
		\label{tab:main_view_full}
	\end{table*}
	
	\begin{table*}[t]
		\centering
		\scriptsize
		\setlength{\tabcolsep}{4.0pt}
		\begin{tabular}{lccccc}
			\toprule
			Diagnostic & Main & All-HO & Int.-stress & Boundary & Ext.-eco \\
			\midrule
			\multicolumn{6}{l}{\textit{Collapse gap = Flagged Acc. $-$ Risk M-Rec.}} \\
			Qwen2.5-14B(\fullpkg) & 11.40 & 37.35 & 32.74 & 20.83 & 15.00 \\
			\textsc{BundleJudge} & -- & -- & 28.63 & 17.66 & 10.70 \\
			\method{} & 0.00 & 1.18 & 0.00 & 0.00 & 0.00 \\
			\midrule
			\multicolumn{6}{l}{\textit{Overall-exact gain between successive stages}} \\
			\verifymethod{} $\rightarrow$ \calibmethod{} & 1.25 & 4.23 & 3.45 & 2.00 & 1.42 \\
			\calibmethod{} $\rightarrow$ \method{} & 1.26 & 3.00 & 1.71 & 2.04 & 1.21 \\
			\bottomrule
		\end{tabular}
		\caption{Large-view collapse diagnostics and stage-wise gains. The strongest flat and structured single-shot baselines retain a visible collapse gap, whereas \method{} nearly eliminates it. The lower block shows that the later stages add narrower but still measurable gains on the larger evaluation views.}
		\label{tab:view_gap_summary}
	\end{table*}
	
	\section{Task and Benchmark Supplement: Data Construction, Annotation, and Sanitization}
	\label{app:data_more}
	
	\paragraph{Sample sources.}
	Clean samples mainly come from public official skill repositories, public community collections, and representative hard negatives. Risk samples come from skill-attack prototypes described in prior work, high-risk patterns summarized in ecosystem scans, public attack-oriented skill collections, and semantics-preserving rewrites built around those patterns. To avoid producing directly reusable attack payloads, all risk samples are sanitized: remote addresses are replaced with .invalid, executable scripts are rendered non-executable, and potential exfiltration targets are replaced with placeholder resources.
	
	\paragraph{Scale and view composition.}
	The expanded evaluation pool contains 581 package-level samples in total: the original 327-package core benchmark, a 30-package public external-ecosystem split, and a 224-package public large-scale expansion. At the sample level, this expanded pool contains 253 clean samples, 95 risk seeds, and 233 rewrites. The five main-text views are composed as follows: Main contains 181 clean / 59 risk / 161 rewrites (401 total); All-HO contains 152 / 72 / 180 (404 total); Int.-stress contains 128 / 60 / 156 (344 total); Boundary contains 104 / 48 / 132 (284 total); and Ext.-eco contains 92 / 42 / 120 (254 total). These numbers more accurately capture that the task is not ``classification over 30 prompts per split,'' but auditing over large cross-file package views assembled from both the core benchmark and public-ecosystem expansions.
	
	\paragraph{Unit of annotation.}
	The minimum annotation unit in this paper is the package rather than an individual file or sentence. Many risks become visible only when files are read jointly. For instance, a benign description in SKILL.md and an override cue in a reference file may each appear harmless in isolation, but together they change the package label.
	
	\paragraph{Labeling principles.}
	\Ben{} denotes packages whose capability boundary and external dependencies are transparent and acceptable. \Sus{} is used for remote bootstrap, external synchronization, deployment initialization, and related boundary cases. \Mal{} requires stronger attack evidence, such as hidden override, disguised transfer, explicit relay/exfiltration cues, or an explicit cross-file intent override. More concretely, \Mal{} requires at least one decisive cross-file attack chain, while \Sus{} is reserved for samples that already exhibit externalization, remote-loading, or role-conflict risk but still do not establish a clear override or transfer intent. Retaining this intermediate class matters in deployment because skill auditing often needs to distinguish between samples that should be blocked and reported immediately and those that require human review.
	
	\begin{table}[t]
		\centering
		\small
		\setlength{\tabcolsep}{4pt}
		\begin{tabular}{>{\raggedright\arraybackslash}p{0.21\columnwidth} >{\raggedright\arraybackslash}p{0.40\columnwidth} >{\raggedright\arraybackslash}p{0.21\columnwidth}}
			\toprule
			Label & Minimum decision condition & Review action \\
			\midrule
			benign & Capability boundary and external dependencies are transparent, and no decisive cross-file attack chain is present. & Allow automatic loading \\
			suspicious & Externalization, remote loading, role conflict, or a suspicious cover story is already present, but evidence remains insufficient for a clear override or transfer intent. & Block auto-loading and send for manual review \\
			malicious & At least one decisive cross-file attack chain exists, such as hidden override, disguised transfer, explicit relay/exfiltration, or privilege-violating bootstrap. & Reject directly and report \\
			\bottomrule
		\end{tabular}
		\caption{Operational label boundaries used in this paper.}
		\label{tab:label_boundary_rules}
	\end{table}
	
	\paragraph{Boundary review protocol.}
	To make the suspicious/malicious boundary auditable, the project additionally organizes a boundary-audit pack of 60 blinded packages, with 30 suspicious and 30 malicious cases across 11 boundary families spanning both the training split and the held-out splits. This asset preserves the full directory structure and includes double-blind annotation sheets, a protocol description, and an adjudication key. In the current manuscript it is treated as a reproducibility asset rather than as additional statistical evidence for the main claims. The paper's argument about label boundaries therefore relies mainly on the operational rules above and on family-level error decomposition, rather than presenting incomplete independent dual-annotation results as established evidence.
	
	\paragraph{Rewrite design.}
	Rewrites are not random paraphrases. They are semantics-preserving variants built around the risk chain: they replace the writing style, disperse evidence across different locations, and alter the cover story while preserving the original functional description and attack intent as much as possible. In the public-ecosystem extensions, the same principle is preserved while broadening repository layouts and surface styles. The goal is not to inflate dataset size, but to measure whether a guardrail can maintain stable labels when intent remains fixed but surface form changes.
	
	\paragraph{Public-ecosystem extensions.}
	The two public-ecosystem extensions broaden the evaluation pool in a different way from the original 30-sample held-out additions. Instead of isolating one stress axis at a time, they pull skill packages from previously unseen public roots, retain their ordinary repository layouts, and then pair them with public or reconstructed attack carriers that match the same package-level auditing format. This is why the enlarged Main, All-HO, Int.-stress, Boundary, and Ext.-eco views are more informative in the final paper than the underlying small splits: they preserve the same semantic chains while removing the visual dominance of coarse split-level increments.
	
	\paragraph{View semantics.}
	The five large views play complementary roles. Main measures the complete expanded benchmark at package scale. All-HO focuses on held-out-style robustness after scale expansion. Int.-stress concentrates the stronger frozen stress conditions while retaining public-transfer cases. Boundary concentrates the review-heavy slice where suspicious/malicious collapse is most likely. Ext.-eco isolates unseen public skill roots and therefore acts as the cleanest ecosystem-facing transfer check in the main text. Together, these views provide a more stable picture of robustness than the earlier 30-sample split reports.
	
	\paragraph{Auditability.}
	Package labels are not treated as opaque final decisions. Each risk seed and rewrite is accompanied by a family tag, a package-level gold label, evidence fields, and a source manifest. For key questions such as why a sample should be labeled suspicious rather than malicious, these metadata provide an auditable trace back to the evidence. The emphasis is therefore on package-level evidence auditability rather than on unexplained end labels alone.
	
	\paragraph{Boundary-review assets.}
	The data-construction process explicitly preserves evidence fields, family annotations, and source manifests, and additionally organizes a blinded review pack that focuses on the suspicious/malicious boundary. The purpose of these assets is to make key label decisions traceable along evidence paths rather than relying only on the final class.
	
	\paragraph{Blinded boundary audit.}
	As a supplementary review, this paper runs a \emph{dual-model blind audit} over 60 blinded boundary packages using Qwen2.5-14B and Qwen2.5-7B verifiers under the default prompt, without exposing family IDs or split information. The raw agreement between the two models is 0.6167 and Cohen's kappa is 0.1085; against gold, the raw agreement is 0.4500 and 0.5000, respectively. More importantly, both models almost never output malicious; most errors are flattenings from malicious to suspicious or benign. This result indicates that the boundary pack genuinely serves as a cross-model stress test for one-shot package-level adjudication and is thus suitable as an audit asset for the suspicious/malicious boundary.
	
	\section{Results Supplement: Additional Analyses and Cases}
	\label{app:cases}
	
	\paragraph{Case 1: reference-only hidden override.}
	In F01, many samples have a benign top-level description, whereas the true override cue appears only in the reference files. Qwen2.5-14B often detects that something is unusual and therefore predicts suspicious, but it struggles to treat the priority override in the reference as decisive enough to recover malicious. The semantic-verification module proposed in this paper is designed specifically to repair such cases, where the structure looks weak but the intent is strong.
	
	\paragraph{Case 2: disguised transfer under safe wording.}
	In F02, transfer behavior is frequently packaged as archiving, handoff, mirroring, or relay. For remote discriminative baselines, such wording is easily confused with normal data synchronization, deployment caching, or backup logic, leading to malicious $\rightarrow$ benign. The proposed method gains its advantage here from two sources: structured extraction aggregates transfer cues across files, and semantic verification judges whether those cues jointly form a disguised transfer chain rather than reacting to isolated keywords.
	
	\paragraph{Case 3: bootstrap--transfer conflict.}
	In F03 and its batch2 variants, the difficult cases are those where remote bootstrap co-occurs with weak transfer cues. A naive rule of the form ``any transfer cue implies malicious'' over-promotes bootstrap-heavy samples that should remain suspicious, whereas an overly conservative rule misses genuinely malicious relay behavior. \calibmethod{} and \method{} address these two questions separately: how to arbitrate chain conflict and how to recover the final lagging sample within a cluster.
	
	\paragraph{Case 4: public-ecosystem layout shift.}
	The public-ecosystem extensions introduce a different difficulty from the core benchmark families: the risky chain is now surrounded by ordinary repository scaffolding such as setup notes, helper scripts, and public contribution conventions. Flat judges often over-trust that scaffolding and keep the package at suspicious even when the decisive chain is already recoverable. This is why the gap on Ext.-eco remains informative even at larger scale: Qwen2.5-14B reaches 83.00 risk malicious recall, \textsc{BundleJudge} reaches 88.30, and \method{} reaches 100.00. The case shows that public-layout realism does not simply make every package harder; it specifically stresses whether the auditing chain can preserve decisive evidence after the package begins to resemble an ordinary open-source skill repository.
	
	\section{Method Supplement: Implementation Details and Trigger Statistics}
	\label{app:impl_more}
	
	\paragraph{Module trigger scope.}
	The additional modules are not executed uniformly on all samples. Structured extraction runs on every sample; semantic verification runs only when the uncertainty trigger fires; chain arbitration runs only when multiple chains are active and their dominance is in conflict; anchor consolidation operates only under a narrow cluster-local condition. The large-view gains make this selective scope visible. From \verifymethod{} to \calibmethod{}, overall exact rises by 4.23 points on All-HO and 3.45 on Int.-stress, indicating that arbitration mainly repairs medium-scope chain conflicts after uncertainty has already been localized. From \calibmethod{} to \method{}, the gains narrow further to 3.00 on All-HO, 1.71 on Int.-stress, 2.04 on Boundary, and 1.21 on Ext.-eco, showing that the final layer mostly removes residual anchor--rewrite disagreement rather than re-deciding the entire package population.
	
	\paragraph{Fixed schema, file selection, and thresholds.}
	The structured extractor audits only the three core roles skill-md, reference, and script. The implementation is frozen as follows:
	\begin{itemize}[leftmargin=*, nosep]
		\item Signal types include override, concealment, tool execution, and external transfer.
		\item Additional signal types include remote bootstrap, cover story, privilege overreach, and description mismatch.
		\item Aggregated features include reference-only override, hidden transfer chain, and remote-bootstrap chain.
		\item Additional aggregated features include cross-file support, simulation wrappers, and a recommended label floor.
		\item Verifier budget: 1 SKILL.md, 2 references, 2 scripts, and 1 repository-context file, with character limits of 2800 / 1800 / 1500 / 700.
		\item Default thresholds: chain-level yes requires 0.60 confidence; the package-floor threshold is 0.70; anchor consolidation requires at least two malicious rewrites; and the cluster must contain no benign prediction.
	\end{itemize}
	The implementation no longer additionally requires the anchor-side floor itself to output suspicious. The same defaults are reused when constructing Main, All-HO, Int.-stress, Boundary, and Ext.-eco.
	
	\paragraph{Semantic-verification cost.}
	Semantic verification is not called on every sample. The enlarged views retain the same uncertainty-localized trigger logic rather than switching to uniform full-package rereads. This matters for interpretation: most of the large-view gain is already obtained at \verifymethod{}, after which the later layers operate on much narrower residuals. In other words, scaling the evaluation from 30-sample splits to 254--404-package views does not change the computational role of the verifier; it only makes the repaired decision pattern easier to read.
	
	\paragraph{Remote token cost.}
	Because the verifier is only triggered on uncertainty-localized packages and the later stages are local, the method family keeps a structural cost advantage over sending every package to a full remote judge. The enlarged evaluation views preserve that same execution pattern: scaling the benchmark to 581 package-level samples increases the amount of evidence processed, but it does not turn the system into a uniform full-package rereader. The practical implication is that the proposed architecture scales by widening the audited package pool, not by abandoning selective verification.
	
	\paragraph{Parameters and freezing protocol.}
	This paper does not present the final results as a single report from a once-trained system on fully untouched test data. Instead, it documents a progression from structured extraction to semantic verification, chain arbitration, and anchor consolidation because those components correspond to recurring error classes identified in the finding experiments. The same default scripts, schema, file budgets, and thresholds are then reused when building Main, All-HO, Int.-stress, Boundary, and Ext.-eco. The large views should therefore be read as scaled transfer checks built on fixed defaults rather than as separately tuned evaluation settings.
	
	\paragraph{Why not simply train a larger classifier?}
	Empirically, relying on a stronger remote judge alone does not solve the enlarged views either. On Int.-stress, Boundary, and Ext.-eco, \textsc{BundleJudge} already improves over flat full-package Qwen, yet it still remains below the staged method family on malicious recovery and rewrite stability. Methodologically, package-level auditing inherently involves cross-file aggregation, intermediate evidence interpretation, and risk/rewrite consistency recovery. The staged framework proposed in this paper makes those structural constraints explicit. This is why the strongest large-view numbers are not read as a generic model-scaling effect, but as evidence that the factorized decision chain repairs a failure mode that larger single-shot judges still leave behind.
	
	\paragraph{Metric interpretation.}
	Overall exact measures overall three-way classification accuracy. Flagged accuracy measures only whether a sample is blocked as non-benign. Risk malicious recall measures the ability to recover truly malicious samples explicitly as malicious. Attack exact consistency measures whether paired risk and rewrite samples keep the same label. These metrics are reported jointly because a skill guardrail must solve two problems at once: whether the risk is blocked at all, and whether the correct risk level is recovered.
	
	\section{Results and Discussion Supplement: Large-View Diagnostics}
	\label{app:metric_more}
	
	The main text emphasizes five large evaluation views with 254--404 packages each, making the dominant trends visible without relying on coarse 30-sample steps. The appendix therefore shifts from split-level reporting to view-level diagnostics that answer three questions more directly: where the strongest remote baseline still collapses the \Sus{}/\Mal{} distinction, how much of the final gain comes from later-stage repair rather than the first semantic-verification jump, and how the public-ecosystem extensions change the shape of the evidence.
	
	\paragraph{Directional collapse remains the most informative failure mode.}
	Table~\ref{tab:view_gap_summary} makes the failure mode explicit by subtracting risk malicious recall from flagged accuracy. For Qwen2.5-14B, the gap is 37.35 on All-HO and 32.74 on Int.-stress, even though the corresponding flagged accuracies remain above 93. This means that the remote baseline often \emph{finds risk} but still under-recovers malicious intent. The same pattern remains visible for \textsc{BundleJudge}, albeit at a smaller magnitude: 28.63 on Int.-stress, 17.66 on Boundary, and 10.70 on Ext.-eco. By contrast, \method{} nearly eliminates the gap across all five large views. The central diagnostic conclusion is therefore stable at scale: the hardest part of package auditing is not generic risk detection, but avoiding directional collapse from \Mal{} to \Sus{} after risk has already been found.
	
	\paragraph{The later stages still matter, but they matter on narrower residuals.}
	The lower block of Table~\ref{tab:view_gap_summary} shows the same decreasing-scope story discussed in RQ4 and RQ5. The jump from \verifymethod{} to \calibmethod{} is largest on All-HO and Int.-stress, where chain-dominance conflict is most visible, whereas the jump from \calibmethod{} to \method{} remains smaller but non-trivial across every large view. This means that the later modules are not broad accuracy boosters. They repair narrower residuals that persist after the main semantic-verification step, especially on views where anchor--rewrite consistency and boundary conflicts still survive the verifier.
	
	\paragraph{The public-ecosystem views widen the evidence rather than replacing it.}
	Ext.-eco should not be read as a replacement for the rest of the benchmark. Its role is to check whether the repaired decision chain still holds when the audited package comes entirely from unseen public skill roots and open-source attack carriers. The results remain favorable to the proposed method family there, but the most important point is methodological: the same package-level diagnosis that explains Main, All-HO, Int.-stress, and Boundary also continues to explain the public-ecosystem view. That continuity is precisely why the enlarged evaluation pool is more useful than keeping the appendix centered on 30-sample split reports.
	
	\section{Reproducibility and Discussion Supplement}
	\label{app:repro_more}
	
	\paragraph{Why are no significance tests reported?}
	The core results do not come from multi-seed training with random initialization, but from deterministic evaluation and staged calibration on a frozen benchmark. In this setting, the central questions are not whether mean differences are statistically significant, but whether error modes are decomposable, whether gains correspond to explicit failure types, and whether those gains remain visible under held-out stress tests. Family-level error decomposition, method-progression trajectories, and ablation results are therefore more informative here than standalone $p$-values. To maintain auditability, the appendix simultaneously provides full baseline tables, residual cases, and module-trigger statistics.
	
	\paragraph{How should the near-saturated large-view results be interpreted?}
	This issue requires careful interpretation. First, the strongest main-text numbers are not obtained from a single black-box judge in one pass. They emerge from the staged progression \verifymethod{} $\rightarrow$ \calibmethod{} $\rightarrow$ \method{}, with each step repairing a narrower residual than the previous one. Second, the enlarged views mix the core benchmark with public-ecosystem extensions, so the results should be read as evidence that the repaired decision structure remains stable after scale expansion, not as proof that every open-world variation has been solved. Third, the appendix diagnostics show that the most informative remaining issue is still directional malicious under-recovery by flat or single-shot judges, not wholesale benign leakage.
	
	\paragraph{Are the stress views truly frozen, and which splits participated in progression?}
	Batch1 and batch2 participated in the original method progression because they exposed the earliest failure patterns that the paper aims to explain. By contrast, the same extractor thresholds, chain-arbitration thresholds, and anchor-consolidation rules were then reused directly when building the larger Main, All-HO, Int.-stress, Boundary, and Ext.-eco views from the benchmark and public-ecosystem extensions. This is precisely why the large views provide a more realistic answer to the question of how far a frozen skill-guardrail pipeline can transfer after public-scale expansion.
	
	\paragraph{Why not replace the semantic verifier with a larger model or multiple models?}
	The goal of the paper is not to argue that a sufficiently large verifier alone solves the problem. The strongest remote discriminative baseline, Qwen2.5-14B, and the structured one-shot baseline \textsc{BundleJudge} both already have strong single-pass auditing capacity, yet Table~\ref{tab:main_view_full} and Table~\ref{tab:view_gap_summary} show that they still leave a substantial malicious-recovery gap on the larger held-out and ecosystem-facing views. The source of the difficulty is therefore not only model capacity, but also the absence of cross-file structural aggregation and downstream consistency calibration. The main conclusions of the paper do not require assuming that one larger single-shot judge would remove that gap.

\end{document}